\documentclass[aps,prb,singlecolumn,superscriptaddress,nobibnotes,citeautoscript,graphicx,amsmath,amssymb]{revtex4-2}

\usepackage[version=4]{mhchem}
\usepackage{siunitx}

\usepackage{amsmath}
\usepackage{mathtools}
\usepackage{amssymb}
\usepackage{bm}
\usepackage{xcolor}
\definecolor{darkblue}{rgb}{0.0,0.0,0.3}
\definecolor{goodblue}{rgb}{0.0,0.0,0.6}
\usepackage{hyperref}

\newcommand{\lef}{\left(}
\newcommand{\rig}{\right)}

\newcommand{\imu}{i}

\newcommand{\deriv}{\mathrm{d}}

\newcommand{\veck}{\mathbf{k}}
\newcommand{\vecr}{\mathbf{r}}

\DeclareMathOperator{\sign}{sign}
\DeclareMathOperator{\Pf}{Pf}

\begin{document}

\title{Robust and fragile Majorana bound states \\ in proximitized topological
	insulator nanoribbons}

\author{Dennis Heffels}
\email[]{d.heffels@fz-juelich.de}
\affiliation{
	Peter Grünberg Institute 9, Forschungszentrum Jülich \& JARA Jülich-Aachen 
	Research Alliance, 52425 Jülich, Germany
}
\affiliation{
	JARA-Institute for Green IT, RWTH Aachen University, 52056 Aachen, Germany
}
\author{Declan Burke}
\author{Malcolm R.\ Connolly}
\affiliation{
	Blackett Laboratory, Imperial College London,
	South Kensington Campus, London SW7 2AZ, United Kingdom
}
\author{Peter Schüffelgen}
\affiliation{
	Peter Grünberg Institute 9, Forschungszentrum Jülich \& JARA Jülich-Aachen 
	Research Alliance, 52425 Jülich, Germany
}
\author{Detlev Grützmacher}
\affiliation{
	Peter Grünberg Institute 9, Forschungszentrum Jülich \& JARA Jülich-Aachen 
	Research Alliance, 52425 Jülich, Germany
}
\affiliation{
	JARA-Institute for Green IT, RWTH Aachen University, 52056 Aachen, Germany
}
\author{Kristof Moors}
\email[]{k.moors@fz-juelich.de}
\affiliation{
	Peter Grünberg Institute 9, Forschungszentrum Jülich \& JARA Jülich-Aachen 
	Research Alliance, 52425 Jülich, Germany
}

\date{\today}

\begin{abstract}
	Topological insulator (TI) nanoribbons with proximity-induced 
  superconductivity are a promising platform for Majorana bound states (MBSs). 
  In this work, we consider a detailed modeling approach for a TI nanoribbon in 
  contact with a superconductor via its top surface, which induces a 
  superconducting gap in its surface-state spectrum. The system displays a rich 
  phase diagram with different numbers of end-localized MBSs as a function of 
  chemical potential and magnetic flux piercing the cross section of the 
  ribbon. These MBSs can be \emph{robust} or \emph{fragile} upon consideration
  of electrostatic disorder. We simulate a tunneling spectroscopy setup to 
  probe the different topological phases of top-proximitized TI nanoribbons. 
  Our simulation results indicate that a \emph{top-proximitized} TI nanoribbon
  is ideally suited for realizing fully gapped topological superconductivity, 
  in particular when the Fermi level is pinned near the Dirac point. In this 
  regime, the setup yields a single pair of MBSs, well separated at opposite 
  ends of the proximitized ribbon, which gives rise to a robust quantized 
  zero-bias conductance peak.
\end{abstract}

\maketitle

\section{\label{sec:introduction}Introduction}

Three-dimensional (3D) topological insulators (TIs) have received a lot of 
attention in the last decade due to their interesting electronic properties, 
in particular due to their topologically protected surface states with a 
spin-momentum-locked Dirac-cone energy spectrum~\cite{Hasan2010}. This 
interest has only increased when the possibility emerged to realize exotic 
forms of superconductivity by combining TIs and ordinary $s$-wave 
superconductors in heterostructures~\cite{Fu2008, Xu2014, Wiedenmann2017, 
Floetotto2018, Schueffelgen2019, Fischer2022, Bai2022}, exploiting the 
superconducting proximity effect~\cite{McMillan1968}. Due to strong 
spin-orbit coupling in the TI, the induced superconductivity transforms into 
$p$-wave pairing for the TI surface states~\cite{Potter2011}.

A promising application of $p$-wave superconductivity is to realize Majorana 
bound states (MBSs) in a {spinless fermionic channel}~\cite{Kitaev2001}, 
forming a so-called Majorana wire. These MBSs come in pairs 
of states at zero energy, which are localized at opposite ends of the 
wire. A pair of MBSs can also be understood as a pair of quasiparticles 
forming equal-weight superpositions of a particle and hole state. As such, 
the MBS is a quasiparticle that is its own anti(quasi)particle, with an 
associated creation/annihilation operator that is 
self-adjoint~\cite{Wilczek2009}.
Furthermore, these MBSs are anyons with non-Abelian exchange 
statistics~\cite{Stern2010}. By combining these exotic properties, MBSs can be 
exploited for quantum information processing with the promise of being immune 
against the most common sources of decoherence~\cite{Freedman2003, Alicea2011, 
Heck2012, Hyart2013, Aasen2016}.

With the abovementioned properties, a superconductor-TI nanowire {or 
nanoribbon} heterostructure appears to be a natural Majorana wire candidate. 
Unfortunately, the surface-state spectrum suffers from a spin degeneracy, 
which naturally arises due to confinement quantization of the 
spin-momentum-locked Dirac cone spectrum with antiperiodic boundary 
conditions for the Dirac spinor solutions~\cite{Zhang2010}. This prevents the 
realization of topological $p$-wave superconductivity with a single spinless 
channel that is underlying the realization of a Majorana wire. The unwanted 
degeneracy can be lifted, however, by applying an external magnetic field along 
the wire~\cite{Ostrovsky2010, Zhang2010, Bardarson2010, Rosenberg2010}. The 
magnetic flux through the cross section of the TI nanowire modifies the 
boundary condition for the surface states and thereby lifts the degeneracy 
that prevents the formation of a topologically nontrivial 
regime~\cite{Cook2011, Cook2012}.

Even when the proximitized TI nanowire is brought into the topological 
regime with an external magnetic field, there is no guarantee that 
well-separated MBSs form at opposite ends of the wire. For this, a sizeable 
proximity-induced superconducting gap must be induced in the surface-state 
spectrum of the TI {nanowire}. In recent works, the 
realization of such a 
fully gapped nontrivial regime was brought into question, as it was found that 
the particle and hole states fail to couple due to a mismatch in transverse 
momentum and, hence, the proximity effect fails to induce a superconducting 
gap~\cite{Juan2019}. To overcome this mismatch, it has been proposed to 
consider a vortex in the superconducting condensate that envelopes the TI 
{wire}~\cite{Juan2019}, or to break the transverse symmetry of 
the {wire} 
with an electric field by introducing an electrostatic gate in the device 
layout~\cite{Legg2021}.

\begin{figure}[tbh]
  \includegraphics[width=0.7\linewidth]{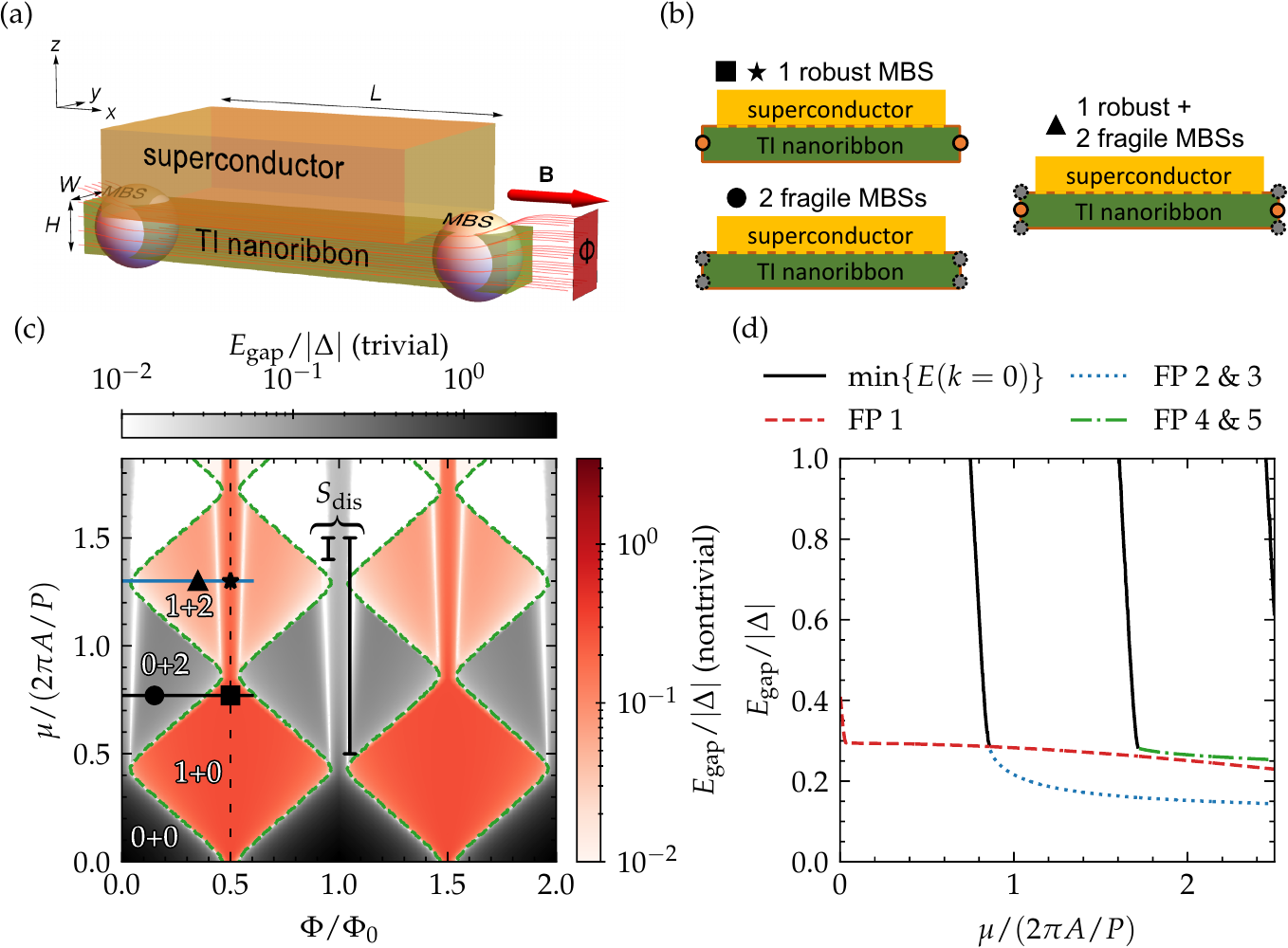}
  \caption{
    (\textbf{a}) A 3D TI nanoribbon that is proximitized by an $s$-wave superconductor 
    via its top surface and pierced by a magnetic flux originating from an 
    external magnetic field oriented along the ribbon. With this setup, MBSs 
    form at opposite ends of the nanoribbon. The setup has different phases 
    with different numbers of robust and fragile MBSs, which are shown 
    schematically in (\textbf{b});
    (\textbf{c}) the spectral gap in the proximitized TI nanoribbon as a function of 
    magnetic flux and chemical potential. The grayscale (redscale) colormap 
    indicates whether the gap is topologically trivial (nontrivial). The phase 
    boundaries between the trivial and nontrivial regions are indicated with 
    green dashed lines. The number of robust plus the number of fragile MBS 
    pairs is indicated for some representative phases;
    (\textbf{d}) a line cut of the spectral gap shown in (\textbf{c}) as a function of chemical 
    potential with magnetic flux equal to half a flux quantum (indicated by a 
    vertical black dotted line in (\textbf{c})). The gap of each surface-state subband 
    that has a Fermi point (FP) when $\Delta = 0$ is presented separately, as 
    well as the gap of the lowest subband without Fermi point, evaluated at $k 
    = 0$.
  }
  \label{fig:1}
\end{figure}

In this article, we scrutinize the conditions for fully gapped topological
superconductivity in a proximitized TI nanoribbon {(i.e., a nanowire with 
rectangular cross section)} structure, considering a 
realistic device layout that is compatible with state-of-the-art 
nanofabrication processes~\cite{Schueffelgen2019}. In particular, we 
consider a selectively grown TI nanoribbon that is covered from the top by a 
conventional superconductor (see Figure~\ref{fig:1}a), e.g., 
Nb~\cite{Schueffelgen2017}. Through careful consideration of the proximity 
effect, we find that this setup naturally yields optimal conditions for fully 
gapped topological superconductivity when the ribbon cross section is pierced 
by (close to) half a magnetic flux quantum, with a single pair of \emph{robust} 
MBSs appearing at the ends of the nanoribbon. For other flux values, we 
identify different gapless and gapped phases with different numbers of 
end-localized MBSs, depending on the position of the Fermi level with respect 
to the Dirac point. Some of these MBSs appear to be \emph{fragile} when 
disorder is introduced in the system and suffer from hybridization with other 
MBS solutions on the same end of the ribbon. We also consider a tunneling 
spectroscopy setup for distinguishing robust and fragile MBSs and identifying 
these different phases in experiments.

Starting with this introduction, the article is divided into six sections. 
In Section~\ref{sec:model}, we cover the details of our simulation approach, 
including the continuum model Hamiltonian (Section \ref{subsec:TI}), the treatment 
of the superconducting proximity effect (Section \ref{subsec:proximity}), and the 
tight-binding modeling approach (Section \ref{subsec:tight-binding_model}). In 
Section~\ref{sec:spectral_gap}, we discuss the (proximity-induced) spectral 
gap and the topologically trivial and nontrivial regimes of a 
top-proximitized TI nanoribbon. In Section~\ref{sec:MBS}, we discuss the 
different phases of this system, with different numbers of robust and 
fragile MBSs, which can be probed with tunneling spectroscopy 
(Section \ref{subsec:tunneling_spectroscopy}). We proceed with a discussion of the 
simulation results in Section~\ref{sec:discussion} and conclude in 
Section~\ref{sec:conclusion}.

\section{\label{sec:model}Model}

\subsection{\label{subsec:TI}Topological Insulator}

For obtaining the TI nanoribbon spectrum, we consider {a tight-binding 
model (see Section~\ref{subsec:tight-binding_model}) that is derived from} the 
following 4-band 
continuum model Hamiltonian for the \ce{Bi2Se3} 
family of 
TI materials~\cite{Zhang2009, Liu2010}:
\begin{equation}
  \label{eq:continuum_model}
  \begin{split}
    H_0(\veck) &= [\epsilon(\veck)-\mu] + M(\veck) \sigma_z
    + A_\perp (k_y s_x - k_x s_y) \sigma_x + A_z k_z \sigma_y \\
    \epsilon(\veck) &\equiv C_0 - C_\perp (k_x^2 + k_y^2) - C_z k_z^2, \qquad
    M(\veck) \equiv M_0 - M_\perp(k_x^2 + k_y^2) - M_z k_z^2.
  \end{split}
\end{equation}
$H_0(\veck)$ is a $4\times 4$ matrix that is written as a linear combination 
of a tensor product of Pauli matrices $s_a$ and $\sigma_b$ ($a, b \in \{x, 
y, z\}$) acting on the spin and orbital (pseudospin) subspaces, 
respectively. Note that the identity matrices are not written explicitly 
here. The parameters $C_0, C_\perp, C_z, M_0, M_\perp, M_z, A_\perp, A_z$ 
can be obtained for different TI materials, such as \ce{Bi2Se3} or 
\ce{Bi2Te3}~\cite{Liu2010}. Here, we will neglect in-plane ($\perp$) versus 
out-of-plane ($z$) anisotropy (in terms of model parameters, $A_\perp = A_z 
\equiv A, C_\perp = C_z \equiv C, M_\perp = M_z \equiv M$) and asymmetry 
between valence and conduction bands ($C = 0$) for simplicity {(these 
simplifications will not significantly affect the surface-state spectrum near 
the Dirac point~\cite{Moors2018}, which is the focus of this work)}. We 
consider remaining model parameters $A = \SI{3}{\eV\cdot\angstrom}$, $M = 
\SI{15}{\eV\cdot\angstrom^2}$, and $M_0 = \SI{0.3}{\eV}$ to represent a 3D TI 
material with inverted (direct) bulk band gap at the $\Gamma$ point equal to 
$0.6\,\textnormal{eV}$ and a Dirac velocity for the 3D TI surface states equal 
to $v_\mathrm{D} = A/\hbar \approx 4.6 \times 10^5\,\textnormal{m/s}${, which are both comparable to \ce{Bi2Se3}~\cite{Zhang2009} (note 
that $0.6\,\textnormal{eV}$ reflects the band separation at the $\Gamma$ point 
and not the overall bulk band gap, which is closer to 
$0.3\,\textnormal{eV}$).}

For describing a charge carrier density fluctuations in the ribbon (i.e., 
electrostatic disorder), we can add a disorder term $S_\mathrm{dis} 
\phi(\mathbf{r})$ to the parameter $C_0$ ($\mu = C_0$ corresponds to the 
chemical potential being pinned at the Dirac point in a pristine system without 
disorder). This disorder is characterized by a fluctuation amplitude 
$S_\mathrm{dis}$ and the function $\phi(\mathbf{r})$, which we consider to be a 
unit-normalized white-noise profile [$\langle \phi( \mathbf{r} ) \phi( 
\mathbf{r}' ) \rangle = \delta ( \mathbf{r} - \mathbf{r}' )$] or a Gaussian 
random field [$\langle \phi( \mathbf{r} ) \phi( \mathbf{r}' ) \rangle = e^{ 
-(\mathbf{r} - \mathbf{r}')^2/(2 \lambda^2)}$] with spatial correlation length 
$\lambda$. Here, we consider an electrostatic disorder strength of the order of 
the TI surface-state subband spacing ($\sim$$2 \pi A / P$, see below, which is 
in the $10\,\text{meV}$ range) and a spatial correlation length in the few-nm 
range~\cite{Chong2020}.

\subsection{\label{subsec:proximity} Proximity-Induced Superconductivity}

In our setup, we consider a 3D~TI nanoribbon that is proximitized via its 
top surface by an $s$-wave superconductor. We make use of the Bogoliubov--de 
Gennes (BdG) formalism to treat the proximity-induced superconducting 
pairing~\cite{Tinkham2004}, which yields the following model Hamiltonian:
\begin{equation}
    \label{eq:H}
    \mathcal{H} = \frac{1}{2} \sum_{\veck} \bm{\Psi}_\mathbf{k}^\dagger
    \underbrace{\left(
    \begin{matrix}
        H_0(\veck) & \underline{\underline{\Delta}} \\
        \underline{\underline{\Delta}}^\dagger & -H^*_0(-\veck)
    \end{matrix}
    \right)}_{ = H_{\mathrm{BdG}}}
    \bm{\Psi}_\mathbf{k},
\end{equation}
with the Nambu spinor $\bm{\Psi}_\mathbf{k}^\dagger = 
(\bm{c}_{\veck}^\dagger, \bm{c}_{-\veck}^T)$ with $\bm{c}_\veck^\dagger = 
(c_{\veck A \uparrow}^\dagger, c_{\veck B \uparrow}^\dagger, c_{\veck A 
\downarrow}^\dagger, c_{\veck B \downarrow}^\dagger)$. Here, 
$c^\dagger_{\veck\alpha}$ ($c_{\veck\alpha}$) are creation (annihilation) 
operators that form a basis for $H_0$ ($-H^*_0$) with degrees of freedom 
$\alpha, \beta \in \{ \uparrow A, \uparrow B, \downarrow A, \downarrow B \}$.

We consider conventional $s$-wave pairing (induced by the superconductor on 
top of the TI nanoribbon) in the BdG formalism, which is given by {a 
momentum-independent pairing}
\begin{equation}
  \label{eq:pairing_potential}
  \underline{\underline{\Delta}} = i \Re\{\Delta\} s_y - \Im\{\Delta\} s_y,
\end{equation}
with complex-valued pairing potential $\Delta = \Re\{\Delta\} + i 
\Im\{\Delta\}$. Combining 
Equations~\eqref{eq:continuum_model}--\eqref{eq:pairing_potential}, we obtain the 
following 8-by-8 BdG-Hamiltonian matrix:
\begin{equation} \label{eq:continuum_model_BdG}
  \begin{split}
    H_{\mathrm{BdG}}(\veck) &= [\epsilon(\veck)-\mu]\tau_z + M(\veck) 
    \tau_z \sigma_z
    + A_\perp ( k_y s_x - k_x \tau_z s_y) \sigma_x + A_z k_z 
    \tau_z \sigma_y \\
    &\hphantom{=} - \Re\{\Delta\} \tau_y s_y - i \Im\{\Delta\} \tau_x s_y,
  \end{split}
\end{equation}
with $\tau_c$ ($c \in \{x, y, z\}$) Pauli matrices acting on the
particle--hole subspace.

Note that the superconducting proximity effect encompasses several aspects, 
such as an induced superconducting gap, an induced pairing potential, and 
induced particle--hole correlations~\cite{Zhu2016}. For our simulation 
approach, based on the BdG Hamiltonian in 
Equation~\eqref{eq:continuum_model_BdG}, we do not explicitly include the 
superconductor on top and only consider the pairing potential that is 
induced {at the interface (see Ref.~\cite{Sitthison2014} for an 
explicit derivation of such a pairing term at the interface)} as input. With this input, we can calculate the 
surface-state quasiparticle spectrum at low energies, with induced 
{spectral (superconducting)} gap, as well as 
particle--hole correlations throughout the complete TI nanoribbon, for 
example~\cite{Chiu2016}.

\subsection{\label{subsec:tight-binding_model}Tight-Binding Model}

For simulating a {proximitized} TI nanoribbon, we discretize the BdG 
Hamiltonian of
Equation~\eqref{eq:continuum_model_BdG} {via the standard procedure} onto a 
regular cubic grid with lattice
constant $a = 1\,\textnormal{nm}$, resulting in a tight-binding model
Hamiltonian with on-site and nearest-neighbor hopping matrices, 
$H_\mathrm{onsite}$ and $H_{\mathrm{hop}\,x,y,z}$, respectively.

For the BdG-Hamiltonian $ H_{\mathrm{BdG}}(\veck)$ described in
Equation~\eqref{eq:continuum_model_BdG}, this results in the following matrices:
\begin{equation} \label{eq:onsite_hopx}
  \begin{split}
    H_{\mathrm{onsite}} &= \lef C_0 -\frac{4C_{\perp} + 2C_z}{a^2} -\mu\rig 
    \tau_z
    + \lef M_0 - \frac{4M_{\perp}+2M_z}{a^2}\rig \tau_z \sigma_z, \\
    &\hphantom{=} - \Re\{\Delta\} \tau_y s_y - i \Im\{\Delta\} \tau_x s_y \\
    H_{\mathrm{hop}\,x} &= \frac{C_{\perp}}{a^2} \tau_z + 
    \frac{M_{\perp}}{a^2} \tau_z \sigma_z
    + \frac{iA_{\perp}}{2a} \tau_z s_y \sigma_x,
  \end{split}
\end{equation}
with comparable hopping matrices along the $y$ and $z$ directions.

{All} the simulation results presented in this work are obtained with this tight-binding model (see \hyperref[sec:simulation_approach]{Appendix} for details on the implementation of the simulation approach and retrieval of the spectral gap). For our setup, we consider a nanoribbon with infinite length $L$ or $L = 1\,\mu\textnormal{m}$ along the $x$-direction, and a square cross section (width $W = 10\,\textnormal{nm}$ along $y$, height $H = 10\,\textnormal{nm}$ along $z$, perimeter $P = 2W + 2H = 40\,\textnormal{nm}$) that is proximitized by an $s$-wave superconductor covering its top surface (see Figure~\ref{fig:1}a).
Because the TI is not an intrinsic superconductor, the pairing potential decays quickly away from the TI-superconductor interface, i.e., over atomic distances~\cite{Chiu2016}. It is therefore reasonable to assume a nonzero $\Delta$ (here, we consider {$\Delta$ to be real, without loss of generality, and equal to $5\,\text{meV}$}) only on the topmost layer of the TI nanoribbon lattice model ($\sim$$1\,\textnormal{nm}$ thick) that is considered to be in direct contact with the superconductor, while $\Delta = 0$ elsewhere in the~lattice.

For the orbital effect of an external magnetic field, we consider the 
Peierls substitution method for the hopping terms:
\begin{equation}
  t_{i \rightarrow j} \rightarrow t_{i \rightarrow j} \exp\lef \imu 
  \frac{q}{\hbar}
  \int\limits_{\vecr_i}^{\vecr_j} \deriv \vecr \cdot \mathbf{A}(\vecr) \rig.
\end{equation}
Here, $t_{i \rightarrow j}$ represents the hopping matrix from site $i$ with
position $\vecr_i$ to $j$ with position $\vecr_j$, which is modified by 
Peierls substitution with a phase depending on the vector potential 
$\mathbf{A}$ with corresponding external magnetic field $\mathbf{B} = 
\vec{\nabla} \times \mathbf{A}$, reduced Planck's constant $\hbar$, and the 
charge $q = \mp e$ (with $e$ the elementary charge) of the charge carrier 
(either a particle or a hole). We consider a constant external magnetic 
field oriented along the nanoribbon (in the frame of reference considered 
here, along the $x$-direction) with $\mathbf{A} = (0, |\mathbf{B}|(z-H), 
0)$. We consider this vector potential such that it vanishes on the topmost 
layer of the nanoribbon ($z = H$) and is compatible with $\mathbf{A} = 
\mathbf{B} = \mathbf{0}$ for $z > H$, avoiding any supercurrent $\propto 
|\Delta| \mathbf{A}$ in our description~\cite{Tinkham2004}. These 
assumptions correspond to a simplified treatment of the experimental setup: 
a vector magnetic field that vanishes completely inside the superconductor 
on top of the TI nanoribbon while neglecting any shielding current.

\section{\label{sec:spectral_gap}Spectral Gap}

In Figure~\ref{fig:1}c, we present the gap in the quasiparticle spectrum of a 
top-proximitized TI nanoribbon as a function of magnetic flux piercing the 
cross section of the ribbon and of chemical potential (note that $\mu = 0$ 
corresponds to the position of the Dirac point of the TI surface-state Dirac 
cone). The parameter space is divided into different gapped regions that are 
separated by gapless phase boundaries or regions. In general, the gap lies 
somewhere between zero and {$|\Delta|$, with $\Delta$} the superconducting pairing 
potential considered on the top surface of the TI nanoribbon (in direct 
contact with the proximitizing superconductor){, which} provides 
a natural upper bound for the proximity-induced superconducting gap. Only near 
$\mu = 0$ and 
integer multiples of the flux quantum does the gap exceed $|\Delta|$. Here, the 
evaluated spectral gap is a trivial insulating gap due to confinement 
quantization $\sim$$\pi A / P$, rather than a proximity-induced 
superconducting gap.

The spectral gap is either topologically trivial or nontrivial, and the 
top-proximitized TI nanoribbon only forms a quantum wire with unpaired MBSs in 
the nontrivial regime. The nature of the gap can be determined with the 
following $\mathbb{Z}_2$ topological invariant~\cite{Kitaev2001},
\begin{align} \label{eq:Pfaffian}
	\mathcal{M} = \sign\{\Pf[H_\mathrm{TINR}(k = 0)] 
	\Pf[H_\mathrm{TINR}(k = \pi/a)]\},
\end{align}
with $\Pf$ short for the Pfaffian and $H_\mathrm{TINR}(k)$ the tight-binding 
model Hamiltonian over the cross section of the TI nanoribbon with wave number 
$k$ along the direction of the ribbon. The trivial and nontrivial regions are 
indicated by color and are in good qualitative agreement with the diamond-tiled 
phase diagram that can be obtained analytically for a cylindrical TI nanowire 
model with $\Delta = 0$~\cite{Cook2011}. A (nonproximitized) cylindrical TI 
nanowire has the following surface-state (particle) spectrum:
\begin{align}
	E_{l}(k) = \pm \hbar v_{\text{D}}
	\sqrt{k^2 + (2 \pi l + 1/2 - \eta)^2/P^2} - \mu,
\end{align}
with $k$ the wave number (for propagation along the nanowire), $l = 0, \pm 
1, \ldots$ the quantum number for quantized angular momentum, $v_\mathrm{D}$ 
the Dirac velocity of the Dirac cone, $\eta = \Phi/\Phi_0$ the total 
magnetic flux piercing the nanoribbon cross section in multiples of flux 
quanta ($\Phi_0 \equiv h/e$), and $P = 2\pi R$ the diameter of the cylindrical 
nanowire with radius $R$. From this expression, it can be seen that the 
spectrum is a subband-quantized Dirac cone that is flux quantum-periodic. By 
evaluating the number of Fermi points $\nu$ with $k > 0$ (or $k < 0$), i.e., 
the number of forward (or backward)-propagating surface states at zero energy 
from the different subbands that cross the chemical potential, the topological 
invariant above can be obtained in an alternative way by evaluating 
$\mathcal{M}=(-1)^{\nu}$. In other words, the system becomes topologically 
nontrivial when there is an odd number of such Fermi points and 
corresponding propagating modes. Each diamond in the phase diagram 
represents a bounded region with a given number of Fermi points, which is 
even for the diamonds in grayscale, and odd for the diamonds in redscale. 
Hence, for a piercing magnetic flux that is a half-integer multiple of one 
flux quantum, the system always has an odd number of Fermi points and 
remains in the nontrivial regime for all values of the chemical potential 
$\mu$ (within the TI nanoribbon bulk gap, as we are only considering the 
topological surface states inside the bulk gap). Conversely, the system is 
always in the trivial regime without an external magnetic field, or when the 
piercing magnetic flux is an integer multiple of $\Phi_0$.

\begin{figure}[htb]
  \includegraphics[width=0.7\linewidth]{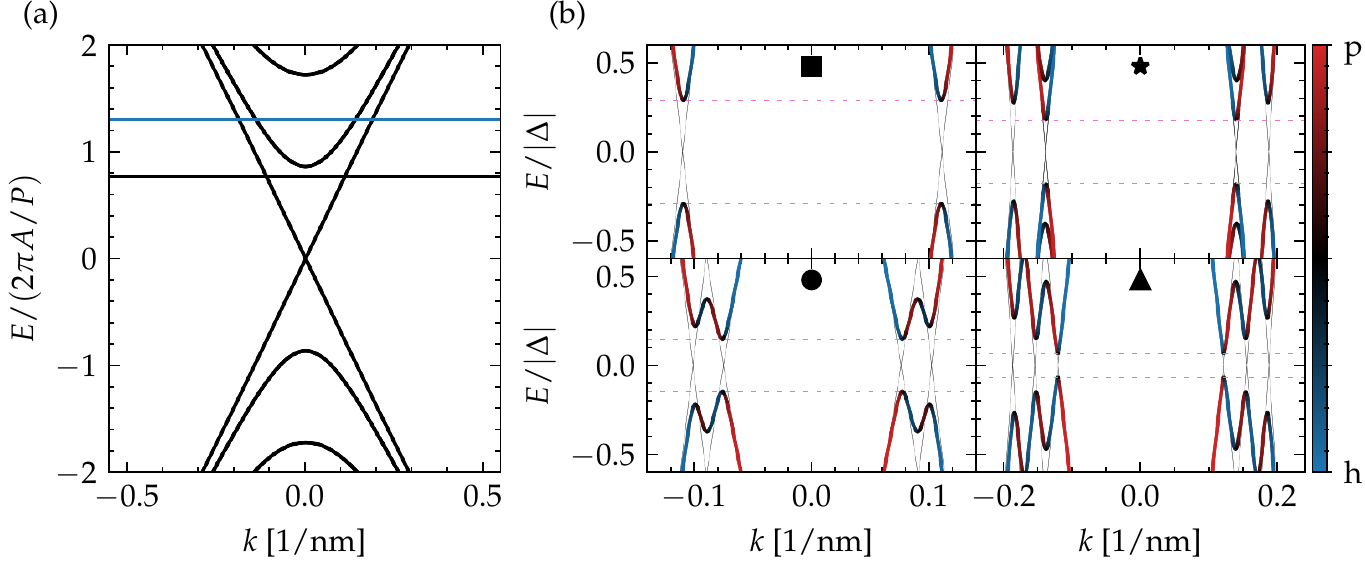}
  \caption{
    {(\textbf{a}) The subband-quantized surface-state spectrum of a TI nanoribbon pierced by half a magnetic flux quantum. The two chemical potential values indicated with line cuts in Figure~\ref{fig:1}c and considered in (\textbf{b}) are indicated with horizontal lines;}
    {(\textbf{b})} the low-energy quasiparticle spectrum of a top-proximitized TI nanoribbon with infinite length for different combinations of piercing magnetic flux and chemical potential, as indicated in Figure~\ref{fig:1}c, with the color indicating the particle--hole (p-h) mixing (blue for hole, red for particle). The spectrum for $\Delta = 0$ is presented with thin black lines and the extent of the spectral gap is indicated with horizontal pink dashed lines.
  }
  \label{fig:1extra}
\end{figure}

Note that Fermi points, i.e., zero-energy surface states, can only be 
considered in general for $\Delta = 0$, as the states can otherwise gap out 
around zero energy. Therefore, we resort to the more general topological 
invariant in Equation~\eqref{eq:Pfaffian} and the perfect diamond tiling gets 
slightly deformed, with different diamonds in the phase diagram becoming 
connected.
Further note that the diamond height (its extent as a function of chemical 
potential) is equal to the subband energy spacing and smaller than $2 \pi A 
/ P$ (with $A = \hbar v_\mathrm{D}$), which is the expected spacing, 
based on the cylindrical wire model when substituting the diameter with the 
perimeter of the square cross section. This {can also be seen in 
the surface-state spectrum presented in Figure~\ref{fig:1extra}a and has been 
reported before}~\cite{Moors2018}{. It may originate from the} pile-up of wave function density near the corners, which 
appears to increase the \emph{effective} perimeter of the cross section.

An interesting finding is that the spectral gap in the nontrivial region is 
maximal for the diamond closest to the Dirac point ($\mu = 0$), which 
corresponds to the region with a single Fermi point, and remains large for the 
diamonds at higher $\mu$ for a piercing flux close to a half-integer flux 
quantum (see Figure~\ref{fig:1}d). The size of this spectral gap agrees well with 
the perturbative estimate $E_\mathrm{gap} \approx \langle \psi_\mathrm{p} \mid 
\underline{\underline{\Delta}} \mid \psi_\mathrm{h} \rangle \sim (W / P) \Delta 
\approx \Delta/4$ for our TI nanoribbon with square cross section, with $\mid 
\psi_\mathrm{p} \rangle$ and $\mid \psi_\mathrm{h} \rangle$ particle and hole 
surface states at the Fermi point with $\Delta = 0$. The reduction factor $1/4$ 
originates from the ratio of the section of the perimeter with nonzero $\Delta$ 
(only the top surface) to the complete perimeter, which gets enveloped by the 
particle and hole surface states.

In addition to the separation into trivial (grayscale) and nontrivial 
(redscale) regions, the spectral gap also reveals different gapped phases within 
the trivial and nontrivial regions themselves{, separated by gapless 
phase boundaries (note that gap closings without a change of topological 
invariant were also reported in Ref.~\cite{Sitthison2014})}. This suggests that 
the regions with and without unpaired MBSs subdivide further into different 
phases with additional distinct properties. As some of the regions stretch out 
over multiple diamonds, we can already rule out that the properties are 
strictly related to the number of Fermi points when $\Delta = 0$.

To reveal the properties of the different phases in the phase diagram, based on 
the spectral gap, we take a closer look at four points (indicated by 
$\blacksquare$, $\bigstar$, $\bullet$, and $\blacktriangle$), which lie in 
different diamonds or regions separated by a gapless boundary in 
Figure~\ref{fig:1}c. Their quasiparticle spectrum is presented in 
Figure~\ref{fig:1extra}{b}. In general, we see the expected number of Fermi 
points 
of the different subbands, based on the diamond to which the point belongs (a 
single Fermi point in the diamond of $\blacksquare$, two Fermi points in the 
diamond of $\bullet$, and three Fermi points in the diamond of $\bigstar$ and 
$\blacktriangle$), and local minima of the proximity-induced superconducting 
quasiparticle gap forming near them. Interestingly, $\bigstar$ belongs to the 
same phase that stretches out over the complete \emph{single-Fermi point} 
diamond below, in which $\blacksquare$ also lies, while $\blacktriangle$ sits 
in the same diamond and represents a different phase, separated from $\bigstar$ 
by a gapless boundary. The qualitative difference between the two phases in the 
quasiparticle can be narrowed down to the number of local minima that can be 
attributed to a single spinless channel. In the diamond with three Fermi points 
where $\bigstar$ and $\blacktriangle$ lie, there is either a single such local 
minimum or three of them, depending on the relative positioning of the Fermi 
points in reciprocal space. For $\bigstar$, two Fermi points overlap in 
momentum, which effectively turns these channels into a trivial spinful 
channel. Hence, the number of spinless channels is reduced to one, which is the 
same numbers as in the diamond below.
This spectral property has consequences for the formation of MBSs, as will 
be discussed in the following section.

\section{\label{sec:MBS}Robust and Fragile Majorana Bound States}

\begin{figure}[htb]
  \includegraphics[width=0.7\linewidth]{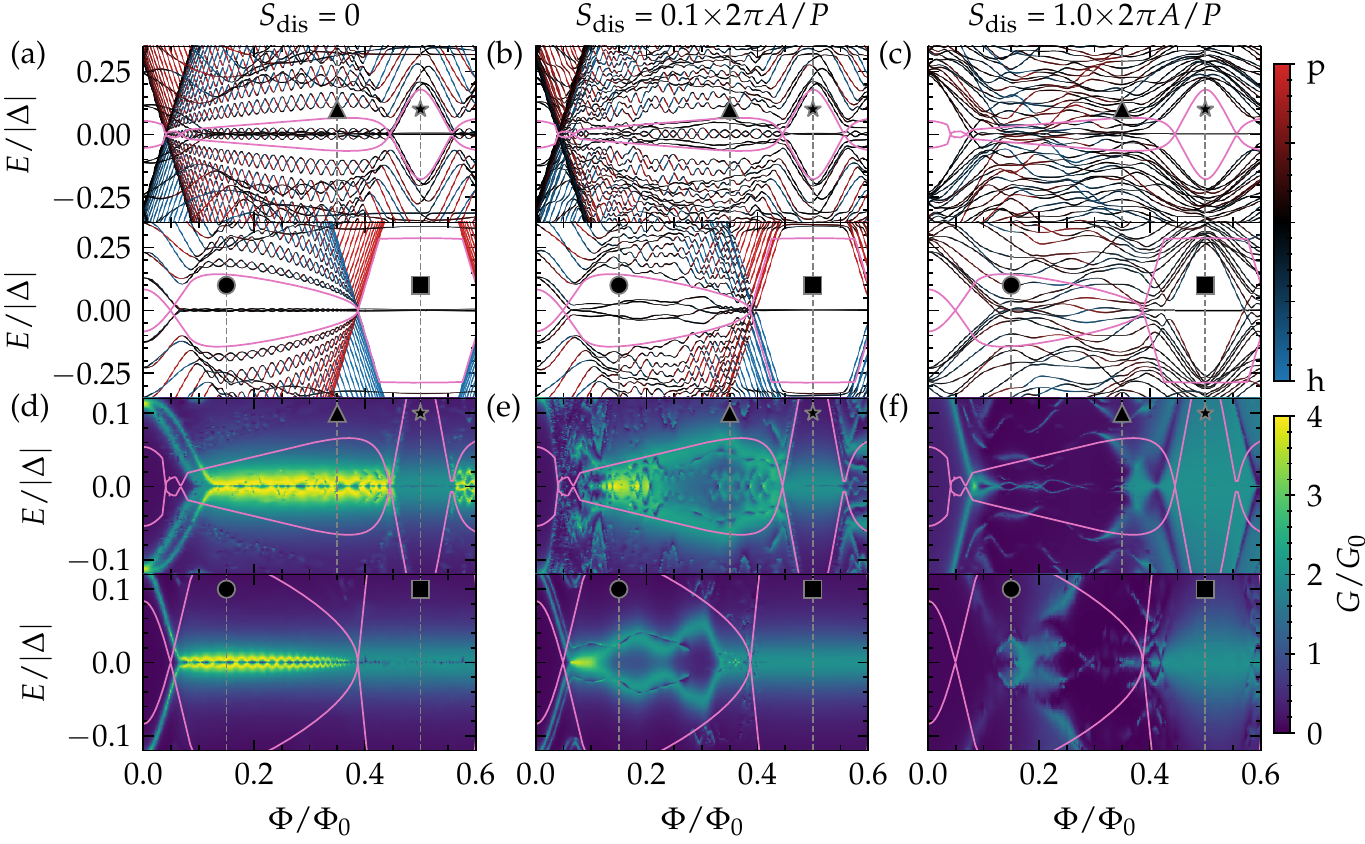}
  \caption{[(\textbf{a}--\textbf{f})] The (\textbf{a}--\textbf{c}) low-energy 
    quasiparticle spectrum of a 
    top-proximitized TI nanoribbon with finite length as a function of piercing 
    magnetic flux $\Phi$, and the {(\textbf{d}--\textbf{f})} tunneling 
    conductance as a function of 
    piercing magnetic flux and energy $E$ (or bias over the tunneling 
    junction).
    The {pink} lines correspond to the spectral 
    gap in the TI nanoribbon with infinite length and without disorder.
    The top (bottom) {panel} of each subfigure corresponds to the 
    upper (lower) 
    horizontal line cut shown in Figure~\ref{fig:1}c at $\mu \approx 1.3 \times 2 
    \pi A / P$ ($\mu \approx 0.75 \times 2 \pi A / P$).
    {The results are obtained (\textbf{a},\textbf{d}) without disorder, and 
    (\textbf{b},\textbf{c},\textbf{e},\textbf{f}) with disorder (considering $S_\mathrm{dis} = 0.1 \times 
    2\pi A/P$ in (\textbf{b},\textbf{e}), and $S_\mathrm{dis} = 2 \pi A / P$ in (\textbf{c},\textbf{f}), with 
    the different disorder strengths also indicated in Figure~\ref{fig:1}c).}
  }
  \label{fig:2}
\end{figure}

\begin{figure}[tb]
  \includegraphics[width=0.7\linewidth]{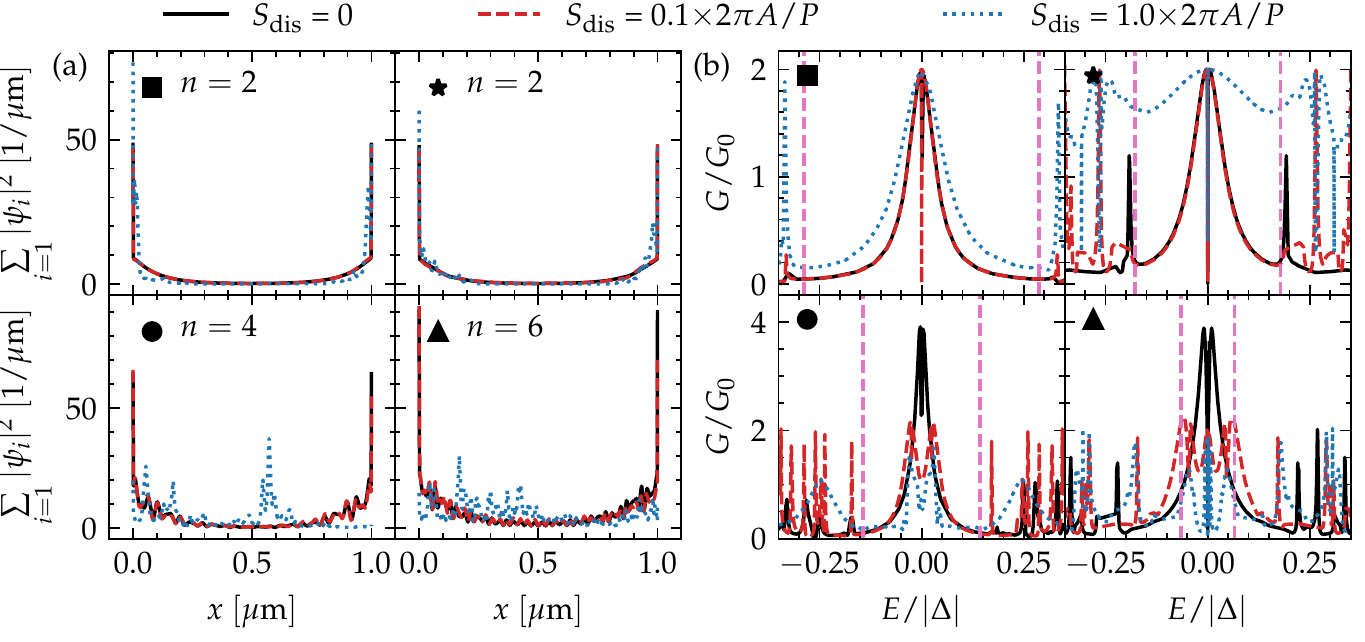}
  \caption{
    [(\textbf{a},\textbf{b})] The (\textbf{a}) one-dimensional wave function density summed over the $n$ lowest-energy (in absolute value) solutions and (\textbf{b}) tunneling conductance as a function of energy of a $1\,\mu$m-long top-proximitized TI nanoribbon with and without disorder with different disorder strengths (indicated in Figure~\ref{fig:1}c) for four different combinations of chemical potential and piercing magnetic flux (also indicated in Figure~\ref{fig:1}c).
    The tunneling conductance in (\textbf{b}) corresponds to the vertical gray 
    dashed line cuts in Figure~\ref{fig:2}d--f. The spectral gap of the TI 
    nanoribbon considering infinite length and no disorder is indicated by 
    vertical pink dashed lines.
  }
  \label{fig:3}
\end{figure}

In Figure~\ref{fig:2}a, we present the low-energy quasiparticle spectrum as a 
function of flux of a top-proximitized TI nanoribbon with finite length. In 
this way, we also reveal the states that are localized at the ends of the 
ribbon. We fix the chemical potential to two different values to explore the 
different trivial and nontrivial phases, as discussed in the section above. 
Near $\Phi = n \Phi_0$ ($n  \in \mathbb{Z}$), there is a completely trivial 
insulating phase without any subgap states. For other values of the flux, 
however, subgap states appear in the spectrum, even in regions that are trivial 
according to the topological invariant in Equation~\eqref{eq:Pfaffian}{, and 
they are localized at both ends of the nanoribbon (see Figure~\ref{fig:3}a)}.

Close to $\Phi = (n + 1/2) \Phi_0$ and for the complete nontrivial diamonds 
nearest to the Dirac point (the connected region containing $\blacksquare$ 
and $\bigstar$, for example), there are only two subgap states. These can be 
identified as a pair of MBSs forming at opposite ends of the top-proximitized 
TI nanoribbon. In other words, this phase corresponds to the conventional 
Majorana quantum wire system. This phase and its MBSs are robust against local 
disorder, as can be seen in {Figure \ref{fig:2}b,c}, 
where it is presented how the low-energy spectra of Figure~\ref{fig:2}a are 
affected by {increasing} electrostatic disorder throughout the nanoribbon.

In the region with $\bullet$, four subgap states form, two on each end of 
the ribbon. In this case, the TI nanoribbon effectively has two independent 
spinless channels (see previous section and Figure~\ref{fig:1extra}), which both 
give rise to the formation of a pair of MBSs forming at opposite ends of the 
nanoribbon. However, as there are two MBSs on each wire end, these MBSs can 
couple to the other MBS on the same end when they are exposed to local 
electrostatic disorder. Hence, in the presence of disorder, these MBSs 
hybridize into non-self charge-conjugate bound states with finite energy. We 
refer to such MBSs as \emph{fragile} MBSs. In contrast, a single unpaired 
MBS can only suffer from hybridization with the MBS on the opposite end of 
the ribbon, which can be suppressed by making the proximitized section 
significantly longer than the MBS localization length $\sim \hbar v_\mathrm{D} 
/ E_\mathrm{gap}$. Therefore, we refer to it as a \emph{robust} MBS pair.
The low-energy spectrum of $\blacktriangle$ shows six subgap states, which 
can be identified as three pairs of MBSs that form at opposite ends of the 
TI nanoribbon. When electrostatic disorder is present, two out of the three 
MBSs can hybridize locally and are thus fragile, while a single robust pair 
should remain protected as long as the TI nanoribbon remains in the nontrivial 
regime (similar to what happens in a tri-junction of Majorana 
wires~\cite{Heck2012}).
{When the electrostatic disorder strength becomes} of the order of 
the diamond height $\sim 2 \pi A / P$, the TI nanoribbon 
is not guaranteed to remain in {in the same phase} when the flux is not an integer ($\Phi = n \Phi_0$) or half-integer 
[$\Phi = (n+1/2)\Phi_0$] multiple of the flux quantum. In this {strongly 
disordered} regime, the spectral gap will fluctuate strongly along the 
disordered nanoribbon and {locally cross gapless phase boundaries. 
Because of this,} many states with near-zero energies {can} appear, which 
renders it difficult to interpret the spectrum in terms of end-localized MBS 
pairs.

Now, we can put all these findings together with the phase diagram and spectral 
properties obtained in the previous section. It becomes clear that the 
different gapped phases can be identified by the number of MBS pairs forming at 
opposite ends of the proximitized TI nanoribbon, with the number always being 
even (odd) in the trivial (nontrivial) regime. When there is some amount of 
electrostatic disorder in the proximitized TI nanoribbon, however, it is 
expected that any even number of MBSs will hybridize locally with the other 
MBSs at the same end, while a single pair of MBSs should remain immune from 
hybridization in a nontrivial regime with an odd number of pairs in total. This 
single pair will then survive near zero energy at opposite ends of the 
proximitized TI nanoribbon.
We can thus identify and label the phases in Figure~\ref{fig:1}c (also see 
Figure~\ref{fig:1}b) by their number of robust MBS pairs (zero or one) and 
fragile MBS pairs (an even number). As the phase with a single robust MBS pair 
and zero fragile pairs stretches out over a large chemical potential window 
near $\Phi = (n+1/2)\Phi_0$, it is the most robust phase with respect to 
electrostatic disorder. It does not suffer from fragile MBS, nor from strong 
fluctuations of the spectral gap.
In the subsection below, we discuss characteristic signatures of these 
different phases and their MBSs in a tunneling spectroscopy setup.

\subsection{\label{subsec:tunneling_spectroscopy}Tunneling Spectroscopy}

In this subsection, we consider the tunneling conductance of a metallic
tunneling probe that is attached {to an uncovered} end of a 
proximitized TI nanoribbon (see \hyperref[sec:simulation_approach]{Appendix} for details), as a function of piercing magnetic flux and energy of the injected carriers from the tunneling probe (corresponding to bias voltage across the tunneling junction).
{Note that, in the experimental setup, the uncovered end should be 
shorter than the induced coherence length $\sim \hbar 
v_\mathrm{D}/E_\mathrm{gap}$~\cite{McMillan1968} for obtaining a \textit{hard} 
proximity-induced gap and clearly revealing the subgap states.} The results of 
these simulations are shown in Figure~\ref{fig:2}{d--f} for the same TI 
nanoribbon and parameters as in Figure~\ref{fig:2}{a--c}, and in 
Figure~\ref{fig:3}b for fixed values of the piercing magnetic flux. We consider 
conductance normalized to $G_0 \equiv e^2/h$.

Overall, we find that the tunneling conductance reveals the subgap 
spectrum of the top-proximitized TI nanoribbon without disorder well, with a 
quantized conductance peak of $2e^2/h$ near zero bias (corresponding to perfect 
Andreev reflection) {when a single unpaired MBS is localized} on the side of 
the tunneling probe. Due to the finite length of the ribbon, there is 
hybridization of MBS pairs across the length of the ribbon, resulting in a 
splitting of the conductance peak away from zero bias. This splitting is 
modulated by the piercing magnetic flux.

When electrostatic disorder is introduced, the tunneling conductance of the 
phase with only a single (robust) MBS pair remains qualitatively the same. The 
zero-bias conductance peak is wider {in the case of strong disorder}, but 
it remains quantized and pinned at zero energy (see Figure~\ref{fig:3}b).
The tunneling conductance of the phases with (fragile) MBS pairs gets heavily 
affected by disorder, however, and a quantized conductance peak cannot be 
easily identified. Instead, the conductance near zero bias 
displays irregular signatures that are very sensitive to the bias and piercing 
flux.
This can be expected as the subgap spectrum itself is heavily affected by the 
disorder. Hence, it will be harder to identify phases with fragile MBSs via 
tunneling spectroscopy, and to determine whether the phase lies in the trivial 
or nontrivial regime{, in particular when the electrostatic disorder 
strength is of the order of the subband spacing or larger}.

\section{\label{sec:discussion}Discussion}

It is important to note here that the realization of fully gapped topological 
superconductivity in proximitized TI nanoribbons has been considered before 
with comparable simulation approaches~\cite{Cook2011, Cook2012, Sitthison2014, 
Juan2019, Legg2021, Legg2022}. It was pointed out by de Juan {et al}. in 
Ref.~\cite{Juan2019} that some form of transverse asymmetry is required to open 
up a superconducting gap in the TI nanoribbon surface-state quasiparticle 
spectrum. Without asymmetry, the induced gap is expected to vanish 
($E_\mathrm{gap} \approx \langle \psi_\mathrm{p} \mid 
\underline{\underline{\Delta}} \mid \psi_\mathrm{h} \rangle \approx 0$) because 
of a mismatch of quantized transverse momentum. Transverse asymmetry can be 
induced by {a superconducting vortex enveloping the 
nanoribbon~\cite{Juan2019}, by electrostatic gating~\cite{Legg2021}, or by 
considering more sophisticated hybrid structures with multiple superconductors 
and gates, which may also enhance the proximity-induced gap away from the Dirac 
point~\cite{Sitthison2014}, for example}. 
Our results, however, suggest that the strong decay of the pairing potential 
away from the intrinsic superconductor is already sufficient to realize the 
required transverse asymmetry when only bringing one of the side surfaces 
(e.g., the top surface) in direct contact with the superconductor. In this way, 
a sizeable proximity-induced gap, $\sim (W/P) |\Delta|$ or $\sim (H/P) 
|\Delta|$, can be naturally achieved near a half-flux quantum of piercing 
magnetic flux{, especially when the Fermi level is close to the Dirac 
point~\cite{Sitthison2014}}. Hence, our findings suggest that a rather 
straightforward device layout (a TI nanoribbon onto which superconducting 
material is deposited) should already be ideally suited for realizing the 
topologically nontrivial regime with well-separated unpaired MBSs.

Furthermore, we note that multiple (fragile) MBSs are not unique to 
top-proximitized TI nanoribbons. They also appear in proximitized 
semiconductor nanowires with a multi-subband treatment~\cite{Stanescu2011}, for 
example. In that case, however, the nontrivial regions with a different number 
of Fermi points are disconnected, such that electrostatic disorder can more 
easily push the system into a trivial regime~\cite{Aghaee2022}.

{For future work, the consideration of material and sample-specific 
parameters for the, e.g., TI model Hamiltonian, ribbon dimensions, and disorder 
strength would be interesting to explore. Equally relevant is the consideration 
of more complicated MBS architectures that allow for braiding, with multiple 
proximitized nanoribbons with different orientations as building blocks (e.g., 
a Y-junction~\cite{Cook2012}). For such structures, the orbital effect from an 
external magnetic field that is misaligned with one of the ribbons should also 
be considered. This has been shown to induce a steering effect in TI nanoribbon 
structures~\cite{Moors2018, Koelzer2021}, and also affects the topological 
phase in semiconductor nanowires, for example~\cite{Lim2012, Nijholt2016, 
Kiczek2017, Winkler2019}.}

Regarding the experimental feasibility, we expect that the phase diagram 
presented here can be resolved in state-of-the-art TI nanoribbon samples. 
Quasi-ballistic transport of topological surface states~\cite{Dufouleur2013, 
	Kim2016} (also in combination with tunability of the Fermi level with respect 
to the Dirac point via electrostatic gating~\cite{Cho2015, Jauregui2016, 
Ziegler2018, Kim2020, Rosenbach2022}), surface-state subband 
quantization~\cite{Muenning2021}, as well as proximity-induced 
superconductivity~\cite{Bai2022}, have all been demonstrated in TI nanowires or 
ribbons. Aside from electrostatic gating, heterostructure engineering can be 
considered to tune the intrinsic Fermi level within a few meV from the Dirac 
point~\cite{Kellner2015, Eschbach2015}.

Finally, we comment on the consideration of tunneling spectroscopy to probe the 
different topological phases of top-proximitized TI nanoribbons. From 
alternative MBS platforms (in particular, semiconductor nanowires), we know 
that a (quantized) zero-bias conductance peak as MBS signature must be 
considered with care, as such a signature can also have a trivial 
origin~\cite{Pan2020}. Nonetheless, the conditions for such false positives are 
less likely to appear in top-proximitized TI nanoribbons because of two 
important differences. First, topological TI surface states have more intrinsic 
robustness against disorder than the low-energy modes of semiconductor 
nanowires due to spin-momentum locking~\cite{Dufouleur2017}. Second, a quantum 
dot at the end of the ribbon, which is one of the important mechanisms for 
retrieving a zero-bias conductance peak with trivial origin in semiconductor 
nanowires~\cite{Pan2020}, is less likely to form because of the 
{linear} Dirac{-cone} spectrum.

\section{\label{sec:conclusion}Conclusion}

With a detailed three-dimensional tight-binding model, we investigate 
numerically the spectral gap of three-dimensional topological insulator 
nanoribbons with magnetic flux-piercing and proximity-induced 
superconductivity, induced by a superconductor on the top surface. The spectral 
gap reveals a rich phase diagram as a function of flux and chemical potential, 
with different gapped phases with paired and unpaired Majorana bound states 
appearing at opposite ends of the nanoribbon. These Majorana bound states can 
be robust or fragile with respect to local hybridization due to electrostatic 
disorder. When the Fermi level in the topological insulator nanoribbon is close 
to the Dirac point and the piercing magnetic flux is close to half a flux 
quantum, we retrieve the optimal conditions for realizing fully gapped 
topological supperconductivity.
With these conditions, there is a single pair of robust MBSs on opposite sides 
of the nanoribbon over an extended range of chemical potential and flux values. 
This phase gives rise to a robust quantized zero-bias conductance peak in 
tunneling spectroscopy.

\begin{acknowledgments}
This work has been supported financially by the German Federal 
Ministry of Education and Research (BMBF) via the Quantum Future project 
'MajoranaChips' (Grant No. 13N15264) within the funding program Photonic 
Research Germany, by Germany's Excellence Strategy Cluster of Excellence 
'Matter and Light for Quantum Computing' (ML4Q) EXC 2004/1–390534769, and by 
the Bavarian Ministry of Economic Affairs, Regional Development and Energy 
within Bavaria's High-Tech Agenda Project "Bausteine für das Quantencomputing 
auf Basis topologischer Materialien mit experimentellen und theoretischen 
Ansätzen" (Grant allocation No. 07 02/686 58/1/21 1/22 2/23).
\end{acknowledgments}

\appendix*
\renewcommand\thefigure{A\arabic{figure}}
\setcounter{figure}{0}

\bibliography{references}

\section{\label{sec:simulation_approach}Simulation Approach}

For the tight-binding simulations, we use the Python package 
Kwant~\cite{Groth2014} with parallel sparse direct solver 
MUMPS~\cite{Amestoy2001} and \emph{Adaptive}~\cite{Nijholt2019} for parameter 
sampling (e.g., flux and chemical potential) on a nonuniform adaptive grid. The 
Pfaffian is calculated numerically with the algorithm of Ref.~\cite{Wimmer2012}.
The source code and raw data of our simulations are archived in a public 
repository~\cite{Juelichrepo}.

\begin{figure}[htb]
  \includegraphics[width=0.4\linewidth]{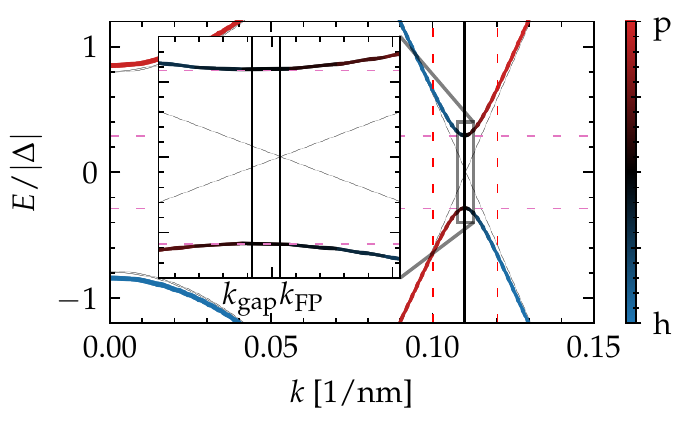}
  \caption{The quasiparticle spectrum of a top-proximitized TI nanoribbon near a local minimum of the gap (indicated by horizontal pink dashed lines), which develops near the Fermi point of the subband crossing of the spectrum with $\Delta = 0$. A narrow interval around each Fermi point (indicated by vertical red dashed lines) is considered to obtain fast convergence towards the value of the spectral gap with an adaptive search algorithm over $k$ values in that interval.}
  \label{fig:A1}
\end{figure}

For the calculation of the spectral gap $E_{\mathrm{gap}}$, we use an 
algorithm that is based on that of Ref.~\cite{Nijholt2016}, but slightly 
modified to speed up the convergence. The modification consists of checking 
whether modes exist for $\Delta=0$ at $E=0$ (i.e., a Fermi point). If not, the 
algorithm from Nijholt {et al}. is used~\cite{Nijholt2016}. If modes exist 
at $E = 0$, we look for a local minimum of the spectral gap as a function of 
$k$ around the Fermi point with an adaptive search 
algorithm~\cite{Nijholt2019}. For $\Delta$ small compared to the subband 
spacing (as is the case in our setup), the difference between the $k$ value of 
the minimum and the subband crossing is small (see Figure~\ref{fig:A1}) so the 
search of this local minimum quickly yields a precise result.

For the tunneling spectroscopy simulations, we consider a metal lead with 
the following continuum model Hamiltonian:
\begin{equation}
  H_\mathrm{metal}(\veck) = \frac{\hbar^2 |\veck|^2}{2 m^\ast} - \mu,
\end{equation}
representing a free electron gas with effective mass $m^\ast = 0.01 \, m_e$ and 
Fermi level $\mu = 20\,\textnormal{eV}$. These artificial model parameters 
allow us to consider a lattice-matched few-channel metallic tunneling probe 
with which the tunneling conductance can be resolved, also when exceeding 
$2e^2/h$ (due to Andreev reflection signatures of multiple MBSs adding up, for 
example).
We discretize this Hamiltonian on the same cubic grid as for the TI nanoribbon 
model, with lattice constant $a = 1\,\textnormal{nm}$, and consider a metal 
lead with translational invariance along $x$ and square cross section of 2-by-2 
lattice sites. 
We attach this lead at one end of a proximitized TI nanoribbon, directly in the 
center of its cross section. We consider a 10-nm-long section of the TI 
nanoribbon that is not covered by a superconductor, separating the metallic 
tunneling probe from the part of the ribbon that is in direct contact with the 
superconductor (with $\Delta \neq 0$ on the top surface).

\end{document}